%% file: main_arxiv.tex
\documentclass[reprint,aps,pra,superscriptaddress,twocolumn]{revtex4-1}

\usepackage{graphicx}

\usepackage{hyperref}

\usepackage{amsmath}
\usepackage{amssymb}
\usepackage{xcolor}

\usepackage{caption}
\usepackage{subcaption}
\input{commands}

\graphicspath{{./figs/}}

\begin{document}


\title{Determination of spatial  quantum states by using Point Diffraction
  Interferometry.}

\author{Quimey Pears Stefano}
\email[]{quimeyps@df.uba.ar}
\affiliation{Universidad de Buenos Aires, Facultad de Ciencias Exactas y Naturales, Departamento de F\'isica, Buenos Aires, Argentina.}
\affiliation{Consejo Nacional de Investigaciones Cient\'ificas y T\'ecnicas, Buenos Aires, Argentina.}

\author{Lorena Reb\'on}
\affiliation{Departamento de F\'isica, IFLP-CONICET, Universidad Nacional de La Plata, C.C. 67, 1900 La Plata, Argentina.}

\author{Claudio Iemmi} \affiliation{Universidad de Buenos Aires,
  Facultad de Ciencias Exactas y Naturales, Departamento de F\'isica,
  Buenos Aires, Argentina.}  \affiliation{Consejo Nacional de
  Investigaciones Cient\'ificas y T\'ecnicas, Buenos Aires,
  Argentina.}

\date{\today}






\begin{abstract}
    \input{abstract.tex}

\end{abstract}
\maketitle
\input{intro.tex}
\input{method.tex}
\input{results_and_conclusions.tex}


\begin{acknowledgments}
This work was supported by Universidad de Buenos Aires (UBACyT Grant
No. 20020170100564BA). %
Q.P.S. was supported by a CONICET Fellowship. 
\end{acknowledgments}

\bibliography{pdiqt-paper}


\end{document}

%% file: commands.tex
\newcommand{\unitMuM}[1]{\ensuremath{#1\,\mathrm{\mu m}}}

\definecolor{color1}{RGB}{102,194,165}
\definecolor{color2}{RGB}{252,141,98}
\definecolor{color3}{RGB}{141,160,203}
\definecolor{color4}{RGB}{231,138,195}

\let\oldhat\hat

\renewcommand{\hat}[1]{\oldhat{\mathbf{#1}}}

%% file: abstract.tex
We present a method to reconstruct pure spatial qudits of arbitrary
dimension $d$, which is based on a point diffraction
interferometer. In the proposed scheme, the quantum states are
codified in the discretized transverse position of a photon field,
once they are sent through an aperture with $d$ slits, and a known
background is added to provide a phase reference. To characterize
these photonic quantum states, the complete phase wavefront is
reconstructed through a phase-shifting technique. Combined with a
multipixel detector, the acquisition can be parallelized, and only
four interferograms are required to reconstruct any pure qudit,
independently of the dimension $d$. We tested the method
experimentally, for reconstructing states of dimension $d=6$ randomly
chosen. A mean fidelity values of $0.95$ is obtained. Additionally, we
develop an experimental scheme that allows to estimate phase
aberrations affecting the wavefront upon propagation, and thus improve
the quantum state estimation. In that regard, we present a
proof-of-principle demonstration that shows the possibility to correct
the influence of turbulence in a free-space communication, recovering
mean fidelity values comparable to the propagation free of turbulence.

%% file: intro.tex
\section{Introduction}
Quantum information processing is a research area in constant growth,
mainly stimulated by promising applications such as: quantum
metrology \cite{Toth2014}, quantum computation \cite{Nielsen}, quantum
networks \cite{Cirac1997}, quantum cryptography \cite{Gisin2002} and
fundamental tests of quantum
mechanics \cite{Ursin2004,Hensen2015}. Therefore, determining the
unknown state of a quantum system is an essential task for the
development of these emerging applications, as well as to compare and
validate their performance. Usually, the schemes for the complete
characterization of a quantum system, collectively named as quantum
state tomography (QST) methods, rely on the result of multiple
measurements on identical copies of the unknown state to estimate its
density matrix $\rho$.  For a quantum system of dimension $d$ (qudit)
there are $d^2-1$ independent elements to be determined in order to
reconstruct $\rho$, posing a scale problem for typical QST methods %
\cite{James2001,Wootters1989,Adamson2010} since they require a number
of measurements that scales as $\sim d^2$.

Despite that, many applications can be improved using high-dimensional
quantum systems
\cite{Cerf2002,Dada2011,Mower2013,Zhong2015,Mirhosseini2015,Martinez2019,Canas2014b}
which pushes increasingly to develop new QST schemes that require
fewer measurements. For example, if some information about the state
is known \emph{a priori}, a reduction in the number of measurements is
feasible. In fact, several works have demonstrated that pure quantum
states can be accurately reconstructed from a number of measurements
that scales as $\sim d$
\cite{Goyeneche2015,Carmeli2016,PearsStefano2019}. Although, in
general, an arbitrary quantum system will be in a mixed state, the
reconstruction of pure states is especially important since most
current applications of quantum information are based on pure states.

Among the several physical implementations of a quantum state,
photonic systems are ideally to be used for quantum communications
applications \cite{Gisin2007}. Particularly, the discretized
transverse momentum of single photons was used to define photonic
quantum states, usually called \emph{slit states}
\cite{Neves2005,Etcheverry2013,PearsStefano2017,Varga2018}. This is a
very versatile option for the encoding of quantum states that allows
to achieve high dimensional Hilbert spaces, easily in relation to
other codifications. In this context, we have previously proposed and
demonstrated \cite{PearsStefano2017} a QST method to reconstruct pure
slit states of dimension $d>2$ based on the phase shifting
interferometry (PSI) technique. As remarkable feature, that scheme
requires a minimum number of measurements, $4d$, and depending on the
used optical architecture, it can be parallelized with the acquisition
of only four interferograms, independently of the system dimension
$d$.

Following the strategy of using interferometric techniques for pure
QST from a reduced number of measurements, we propose here a PSI
scheme based on a point diffraction interferometer (PDI) as a new tool
for high-quality estimation of photonic quantum states.  The PDI
technique, introduced by Linnik in reference \cite{Linnik1933}, uses a
common-path configuration where the reference beam is generated from
the same wavefront under characterization.  Its common-path feature
results on an interferometer extremely stable against vibrations and
air turbulence. It has been applied for testing optical components in
a wide range of wavelengths, from infrared \cite{Koliopoulos1978} to
extreme ultraviolet \cite{Naulleau2000}. When combined with PSI
schemes, for example by using liquid crystal technology
\cite{Mercer1996,Iemmi2003,Ramirez2013} to control the phase steps,
the potential applications of this interferometer are enhanced. For
instance, in reference \cite{Iemmi2003} Iemmi \emph{et al} used a
commercial liquid crystal display (LCD) to implement a PDI scheme that
allows to correct \emph{in situ} the aberrations in a Vander Lugt
correlator, while in reference \cite{Ramirez2013} the technique was
used to obtain a digital holographic movie.

The architecture of the present device is very stable, requires
relatively few optical elements, and it is easy to align. Moreover,
since the detection is performed in the conjugated plane of the input
wavefront, where the slit states are codified, it is possible to
parallelize the measurements and carry out the tomographic process
from only 4 measurements, regardless of the dimension of the quantum
state. To implement this process, it is necessary to modify the
encoding of the slit states, so that, the input signal includes a
coherent background that acts as a reference for the
interferograms. Thus, the technique allows evaluating all points of
the wavefront and not only those belonging to the slits. This feature
makes the method especially suitable for quantum communications in
free space since it allows correcting the effect of turbulence-induced
aberrations.

The article is organized as follows: in section \ref{sec:method} we
review the fundamentals of PDI (\ref{subsec:pdi-fundamentals}) and
propose an alternative encoding for the slit states that includes a
reference beam (\ref{subsec:encoding}). In section \ref{sec:setup} we
describe the experimental common-path interferometer used to perform
the quantum state reconstruction. In section \ref{sec:results},
dedicated to the experimental results, we show the fidelity of
reconstruction for a great number of quantum states. In
  particular, we devote subsection \ref{subsec:turbulence} to show
the feasibility of this method to correct the effects of turbulence in
free-space communication. Finally, in section \ref{sec:conclusion}
we give the conclusions.

%% file: method.tex
\section{Description of the method}\label{sec:method}
In this section we will describe the two main points on which the
tomographic method is based. On one hand, we will briefly review the
features of a PDI adapted for the implementation of PSI techniques,
since it is an accurate and effective method for phase measurement.
On the other hand, we will analyze how the encoding of the slit states
should be modified, with respect to the standard one, in order to be
correctly evaluated with the proposed device.

\begin{figure}[ht]
  \centering
  \includegraphics[width=\linewidth]{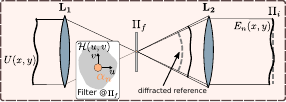}
  \caption{Schematics of the working principle of a PDI. Lens
    $\mathbf{L_1}$ focus the input wavefront to be characterized
    $U(x,y)$ (wavy line) at the Fourier plane, $\Pi_f$. The phase
    filter at $\Pi_f$ introduces a phase $\alpha_n$ in a small area in
    the center of the Fourier transform (see inset) given place to a
    reference wavefront (dashed line). Lens $\mathbf{L_2}$ images
    $U(x, y)$ onto the final plane $\Pi_i$, where the interference
    pattern $|E_n(x,y)|^2$ is registered.}\label{fig:idea}
\end{figure}

\subsection{Point diffraction interferometer (PDI)}
\label{subsec:pdi-fundamentals}

Let us describe the interferometric process following the sketch shown
in figure \ref{fig:idea}.  The input wavefront $U(x, y)$ is focused by
the lens $\mathbf{L_1}$ at the Fourier transform plane $\Pi_f$. When
no diffracting object or aberration is present, the resulting light
distribution is a bright central spot, corresponding to the Fourier
transform of the entrance pupil of the system. At $\Pi_f$, a phase
filter $H(u, v)$, smaller than the focused spot, is placed (see inset
in figure \ref{fig:idea}). This filter is used as a perturbation to
generate a spherical wave by diffraction effect. When an object or
aberration is present in the input signal, the bright spot is
deformed. Most of the light is diffracted towards higher spatial
frequencies, and only the small central part of the spot go through
the phase filter. In this way, after $\Pi_f$ two waves will be
present: an object wave (wavy line in the figure \ref{fig:idea}) and a
reference wave (dashed line), which are going to interfere. Lens
$\mathbf{L_2}$ images the object plane $\Pi_i$ in such a way that the
amplitude and phase distribution at the object plane can be evaluated.

The input wavefront $U(x,y)$ can be fully reconstructed by
implementing a PSI scheme \cite{Creath1988}. This technique consists
in introducing successive controlled phase shifts $\alpha_n$ in the
reference beam that interfere with the object beam, while the
corresponding interferogram intensity is registered.

The transfer function that implements the phase shifts $\alpha_n$ can
be described as
\begin{align}\label{eq:pdi-transfer}
  \mathcal{H}(u,v) = 1 + \delta(u,v) \left[\exp(i \alpha_n) -1\right],
\end{align}
where $\delta(,)$ is the two dimensional Dirac function, $(u,v)$
indicate the transverse coordinates at $\Pi_f$, and
$\alpha_n = 2\pi n/N$ is the constant phase added to the reference in
each of the $N$ steps of a PSI scheme.

The amplitude at the image plane $\Pi_i$ is
\begin{align}\label{eq:en}
  E_n(x,y) = U(x,y)  \circledast h(x,y) = U(x,y) + K \left[ e^{i \alpha_n} - 1\right],
\end{align}
where the symbol $\circledast$ represents the convolution, $h(x,y)$ is
the filter impulse response, and $K$ is the complex constant
corresponding to the mean value of $U(x,y)$. As the final aim of this
method is to obtain the amplitude and the phase of $U(x,y)$, it is
useful to rewrite $U(x,y) = u(x,y) \exp(i \phi(x,y))$, where $u(x,y)$
is the absolute value of $U(x,y)$ and $\phi(x,y)$ represents the
phase.  To fully reconstruct the input wavefront, $N>3$ interferograms
$E_n(x,y)$ ($n=0,\dots,N$), have to be recorded with a multipixel
detector, and the measurements intensities are then combined in the
expressions
\begin{align}\label{eq:C}
  C(x,y) = \sum_{n=0}^{N-1} \left|E_n(x,y)\right|^ 2\cos\left(\frac{2\pi
  n}{N}\right)\\\label{eq:S}
  S(x,y) =  \sum_{n=0}^{N-1} \left|E_n(x,y)\right|^ 2\sin\left(\frac{2\pi n}{N}\right).
\end{align}
Both expression can be simplified according to the orthogonality properties of the trigonometric functions:
\begin{align}\label{eq:C-simp}
  C(x,y)& = -N\left|K\right|^2 + N \left|K\right| u(x,y)
         \cos\left(\phi(x,y) + \mu \right)\\\label{eq:S-simp}
  S(x,y)& =  N \left|K\right| u(x,y) \sin\left(\phi(x,y) + \mu \right).
\end{align}

Thus, the unknown phase of the wavefront $\phi(x,y)$ can be reconstructed
as
\begin{align} \label{eq:phase} \phi(x,y) = \mathrm{arctan2}(S, C-C_0) -
  \mu,
\end{align}
where $\mathrm{arctan2}(x_1, x_0)$ is defined as the angle between the
2-dimensional vector $(x_0 , x_1)$ and the $x_0$ axis, $\mu$ is the
global phase of $K$ and $C_0 = -N\left|K\right|^2$. The value of $C_0$
can be readily obtained from the value of $C(x,y)$ at the points in
which the input wavefront $U(x,y)$ is zero. Finally, the real
amplitude of $u(x,y)$ is $E_0$.

\subsection{Encoding of the spatial photonic qudit}
\label{subsec:encoding}

As we have explained in the previous subsection, the PDI method
requires a nonzero DC component in the input light distribution that
gives place to the reference wave, and for this purpose, we have to
provide a background to the codified states.
\begin{figure}[ht]
  \centering {\includegraphics[width=\linewidth]{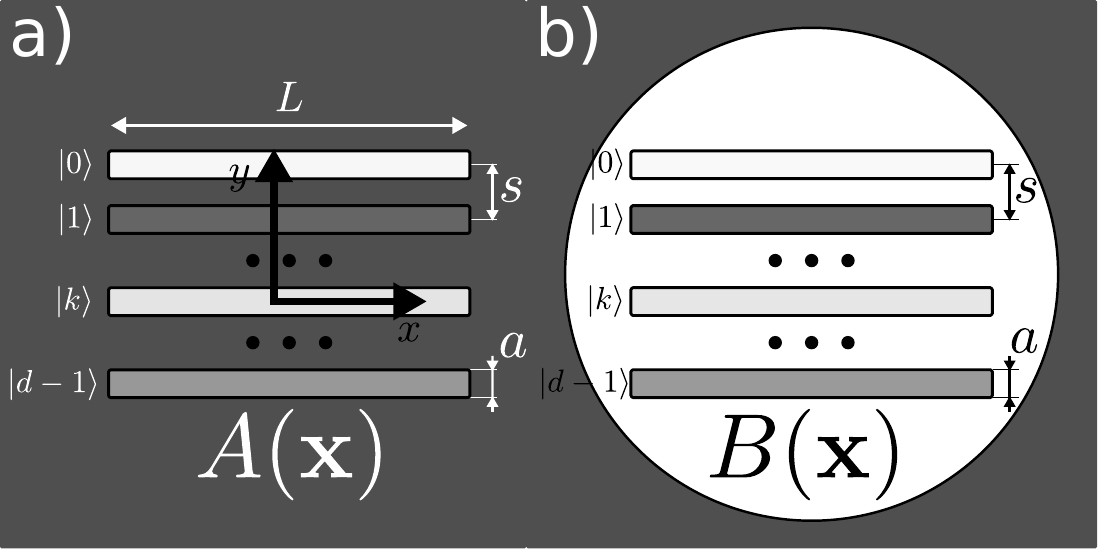}
    \phantomsubcaption\label{fig:slit-mask:a}
    \phantomsubcaption\label{fig:slit-mask:b}}
  \caption{Complex amplitude mask for an example slit state of
    arbitrary dimension $d$. The gray scale is only schematic, and the
    lighter zones represent greater transmissivity. (a) Standard
    definition of a slit state (equation \eqref{eq:slit-mask:a}). (b)
    Proposed modification to the definition of the state in (a). The
    background acts as a reference for the PDI. }\label{fig:slit-mask}
\end{figure}

The formalism of slit states is described in detail in references %
\cite{Neves2004,Solis-Prosser2013,Varga2017}. Here we present a brief
review of the standard codification. Let $A(\mathbf{x})$ be the
complex transmission function of an aperture. When a paraxial and
monochromatic single-photon field, described by the normalized
transverse probability amplitude $\psi(\mathbf{x})$, impinges on this
aperture, the resulting quantum state is
\begin{align}\label{eq:spatial-qudit}
  |\Psi\rangle  = \int \mathrm{d}\mathbf{x} \psi(\mathbf{x})
  A(\mathbf{x}) | 1\mathbf{x}\rangle,
\end{align}
where $\mathbf{x} =(x,y)$ is the transverse position coordinate and
$| 1\mathbf{x}\rangle$ is the single-photon state in the position
basis. We are interested in representing a quantum state of finite
dimension $d$, and therefore, we must discretize the transverse
position. Let us consider that $A(\mathbf{x})$ is an array of $d$
rectangular slits of width $a$, where the separation between adjacent
slits is $s$, and the length is $L (\gg a,s)$, that is:
\begin{align}\label{eq:slit-mask:a}
  A(\mathbf{x}) = \mathrm{Rect}(x/L) \sum_{k=0}^{d-1} c_k \mathrm{Rect}\left(\frac{y-k
  s}{a}\right), 
\end{align}
where the $\{c_k\}_{k=0}^{d-1}$ are the complex transmission
amplitudes of each rectangular region and $\mathrm{Rect}(\eta)$ is the
rectangle function, that takes the value 1 if $|\eta|< 1/2$ and the
value 0 otherwise. Figure \ref{fig:slit-mask:a} shows, schematically,
the aperture given by equation \eqref{eq:slit-mask:a}. For simplicity
we will assume that $\psi(\mathbf{x})$ is approximately constant
across the region of the slits, and hence, the quantum state described
by equation \eqref{eq:spatial-qudit} is the qudit state
\begin{align}\label{eq:qudit}
  |\Psi\rangle = \sum_{k=0}^{d-1} c_k | k\rangle, 
\end{align}
where $\left\{| k\rangle\right\}_{k=0}^{d-1}$ is the logical
basis.

In order to have a light distribution similar to those usually present
in an in-line holography process (small diffracting objects immersed in
a strong background), we propose a modification on the representation
of the slit states that includes a constant light background. These
states can be thought of as generated by an aperture function
$B(\mathbf{x})$ that is the same as $A(\mathbf{x})$ in the zones
corresponding to every slit, but surrounded by a circular region of
constant amplitude that introduces the required background. The
modified aperture, schematically shown in figure
\ref{fig:slit-mask:b},is expressed as
\begin{align}\label{eq:slit-mask-b} B(\mathbf{x}) = &
  A(\mathbf{x}) +\\\nonumber
& \mathrm{Circle}(|\mathbf{x}|^2/R) \left[1 - \mathrm{Rect}(x/L)
\sum_{k=0}^{d-1} \mathrm{Rect}\left(\frac{y-k s}{a}\right) \right],
\end{align} where $\mathrm{Circle}(\eta)=1$ for $|\eta|<1$ and  0
otherwise, and $R$ is the radius of the background pupil.

It is worth remarking that, given a quantum state defined by
$B(\mathbf{x})$, one can always obtain the usual slit state
$|\Psi\rangle$ by postselection with an amplitude mask with unitary
transmission in the slits rectangles, and zero elsewhere.
Additionally, the proposed representation allows to evaluate the
amplitude and phase distribution over all the circular pupil. Later,
in subsection \ref{subsec:turbulence}, we will show how to estimate
the introduced phase aberrations on the photonic quantum state of
interest due to its free propagation in a turbulent medium, and how to
use this information for a corrected estimation of such a state.

\section{Experimental implementation}
\label{sec:setup}

\begin{figure*}[ht]
  \centering
  \includegraphics[width=.95\linewidth]{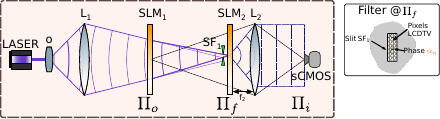}
  \caption{Experimental setup. The light source is a $405\mathrm{nm}$
    cw laser diode, attenuated down to the single photon level. Lenses
    $\mathbf{Ls}$ conform a convergent optical processor.
    $\mathbf{SLMs}$ are phase-only spatial light modulators and
    $\mathbf{SF_1}$ is a spatial filter. The interferograms are
    detected by a high-sensitivity $\mathbf{sCMOS}$ camera. The detail
    shows the filter placed in the Fourier Plane $\Pi_f$: the slit
    $\mathbf{SF_1}$ selects the first diffracted order in the spatial
    qudit preparation, while the central pixel of  $\mathbf{SLM_2}$
    introduces the PSI phase retardation $\alpha_n$.\label{fig:setup}}
\end{figure*}

The experimental setup, depicted in figure \ref{fig:setup}, is
basically a convergent optical processor in which two phase-only
Spatial Light Modulators (SLMs) are used: one to prepare the input
qudit state, and the other to dynamically introduce the phase
retardation needed to implement the PSI process. Both SLMs are
conformed by a Sony liquid crystal television panel model LCX012BL
which, in combination with polarizers and wave plates, that provide
the adequate state of light polarization, allows a $2\pi$ phase
modulation of the incident wavefront 
\cite{Marquez2001}. This model of liquid crystal has a VGA resolution
of $640\times480$ and a pixel size of $\unitMuM{43}$.  The light
source is a laser diode $@405\mathrm{nm}$, that is expanded by the
microscope objective $\mathbf{O}$ and spatially filtered. Neutral
density filters, not shown in the figure, attenuate the source to the
single-photon level. The lens $\mathbf{L}_1$ images the pinhole onto
the Fourier plane $\Pi_f$. The aperture $B(\mathbf{x})$ that defines
the spatial qudit is displayed, dynamically, in the phase-only
$\mathbf{SLM_1}$, which is placed in the object plane
$\Pi_{\mathrm{o}}$. The aperture was programmed so that each slit had
a width $a$ of 4 pixels and a separation $s$ of 6 pixels. To generate
the mask described in equation \eqref{eq:slit-mask-b} by means of a
pure phase modulator we use the method proposed in reference
\cite{Solis-Prosser2013}. Briefly, the mask is encoded as a blazed
phase diffraction grating, where the local phase depth controls the
efficiency of the first diffraction order, and thus the local
amplitude of the mask. The phase of the complex coefficients $c_k$ are
controlled by displacing the grating. If $p$ is the period in pixels,
the relative phase introduced by a displacement of $k$ pixels is
$2\pi \frac{k}{p}$. In our case, the grating period is $p = 12$,
resulting in a sufficiently small discretization error
\cite{Varga2014}. The diffracted first order is selected by means of
the spatial filter $\mathbf{SF_1}$ (a slit of width
$\approx \unitMuM{200}$), which is placed at the $\Pi_f$ plane. The
lens $\mathbf{L_2}$ images the filtered $\Pi_o$ complex distribution
onto the plane $\Pi_i$.

$\mathbf{SLM_2}$ is placed in plane $\Pi_f$, right after the spatial
filter $\mathbf{SF_1}$. We use this modulator to represent the phase
filter $\mathcal{H}(u,v)$ given by equation \eqref{eq:pdi-transfer}),
i.e., the phase shifts $\alpha_n$ needed to implement the PSI
technique are introduced in the central pixel. The inset in figure
\ref{fig:setup} shows a detail of such filters: the pixel that
introduces the phase $\alpha_n$ to the reference (the dashed one
within the pixel structure of the SLM), and the spatial filter
$\mathbf{SF_1}$, that selects the first diffracted order of the
preparation mask displayed in $\mathbf{SLM_1}$.  After
$\mathbf{SLM_2}$, the lens $\mathbf{L_2}$ is placed at a distance equal
to its focal length $f_2$. In this way, the spherical reference
wavefront, which is diffracted by the central pixel, is collimated and
it interferes with the tested wavefront at plane $\Pi_i$. Finally, the
interferograms were detected by a high-sensitivity camera based on
complementary metal-oxide semiconductor ($\mathbf{sCMOS}$) technology
placed at the image plane $\Pi_i$. The camera used is a Thorlabs
Quantalux sCMOS with HD definition.

%% file: results_and_conclusions.tex
\section{Results}
\label{sec:results}
We present here the experimental results that validate our method to
carry out spatial qudit tomography from a minimum number of
measurements. Additionally, in subsection \ref{subsec:turbulence} we
present a proof-of-principle demonstration that the proposed technique
can be used to correct phase aberrations such as the one that
undergoes the free propagation of photonic states under turbulence.

\subsection{Reconstruction of spatial qudits}\label{subsec:results}
\begin{figure}[ht]
  \centering
  {\includegraphics[width=\linewidth]{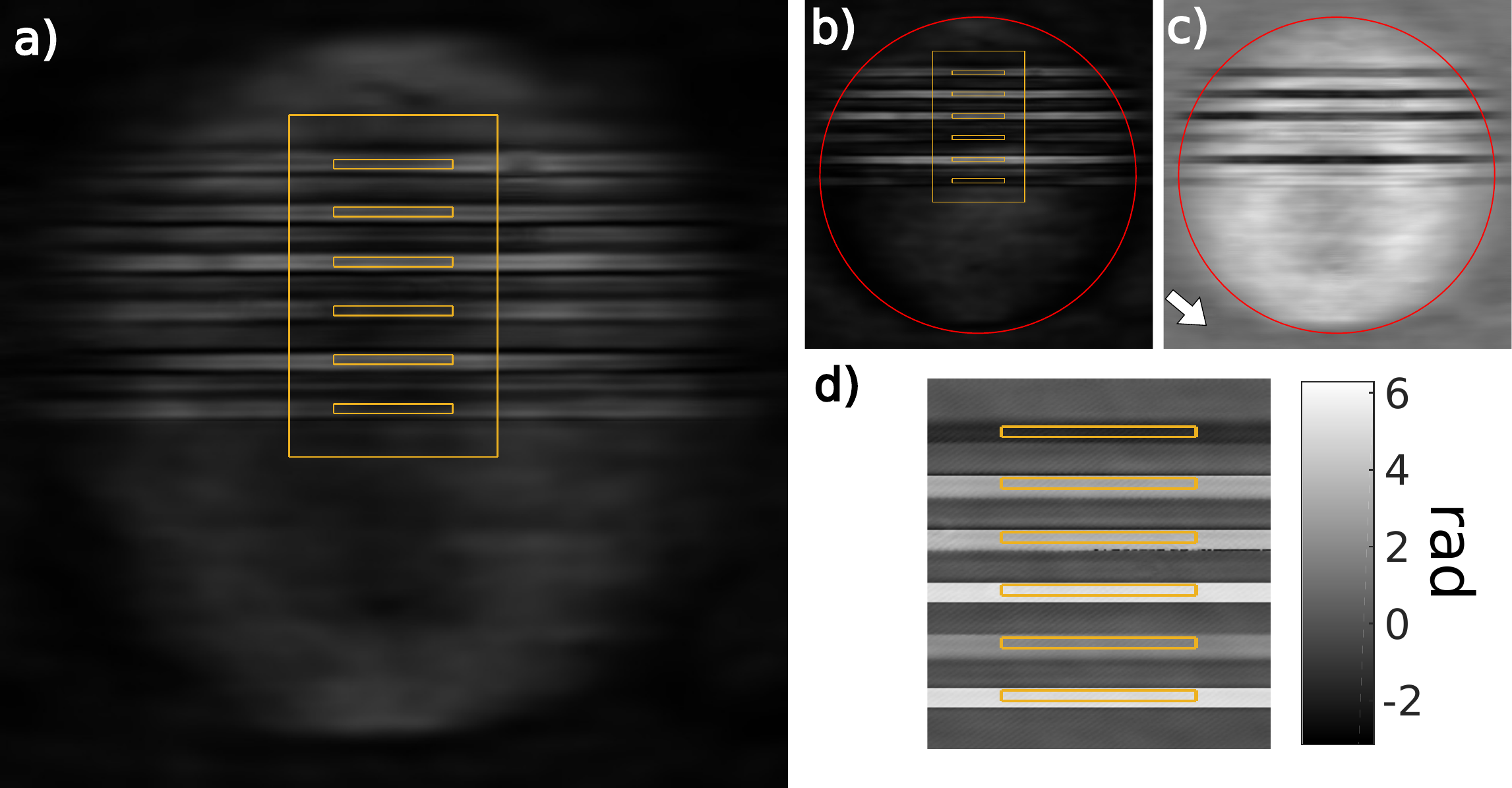}
  \phantomsubcaption\label{fig:interf:a}
  \phantomsubcaption\label{fig:interf:b}
  \phantomsubcaption\label{fig:interf:c}
  \phantomsubcaption\label{fig:interf:d}}
\caption{Example of a measurement for a qudit of dimension $d = 6$
  with uniform amplitudes. (a) Interferogram corresponding to a
  phase shift $\alpha_0=0$, which matches the intensity distribution
  of the aperture $B(\mathbf{x})$. The inner yellow rectangles
  represent the ROIs where the phase of each slit is evaluated. The
  outer rectangle demarcates the zone zoomed in (d). (b)
  Interferogram corresponding to a phase shift $\alpha_2=\pi$. The red
  circle represents the pupil of radius $R$. (c) 2D map of
  corresponding to the term $C(x,y)$. The white arrow points the area,
  outside the pupil, in which $|K|$ is evaluated. (d)
  Reconstructed phase map, the value assigned to each slit is obtained
  by averaging within each ROI. \label{fig:interf}}
\end{figure}

In figure \ref{fig:interf} we show, as an example, the interferograms
registered for a particular qudit state of dimension $d=6$ with
uniform real amplitudes, i.e., where the coefficients $c_k$ in
equation \eqref{eq:qudit} meet that $\left|c_k\right|^2 = 1/d$, and
its phases were arbitrarily selected. The full characterization of the
state is obtained from a four step PSI, $N=4$.  Figure
\ref{fig:interf:a} shows the intensity distribution registered on
$\Pi_i$. The rectangles in yellow indicate the regions of interest
(ROIs) from where the phase and intensity of each slit are
evaluated. It is worth noting that the dark zones surrounding the
slits are the result of diffraction in the borders, but inside the
ROIs both, amplitude and phase of the coefficients $c_k$, are
correctly represented. Figure \ref{fig:interf:b} shows the
interferogram corresponding to a phase shift $\alpha_2 = \pi$, while
the red circle indicates the circular pupil of radius R mathematically
described by the circle function in equation
\eqref{eq:slit-mask-b}. The light intensity \emph{outside} the pupil
in this interferogram corresponds to the reference beam. As it was
explained in \ref{subsec:pdi-fundamentals}, in order to estimate the
reference amplitude $|K|$ we have to know the magnitude of $C(x,y)$,
shown in figure \ref{fig:interf:c}, at those points where the input
wavefront is zero. The white arrow marks an area in which $C_0$ can be
estimated. Finally, figure \ref{fig:interf:d} shows the phase map
corresponding to the zone delimited by the outer yellow rectangle
drawn in figure \ref{fig:interf:a}. The phase value assigned to each
slit is obtained by averaging within each ROI.

\begin{figure}[ht]
  \centering
  \includegraphics[width=.695\linewidth]{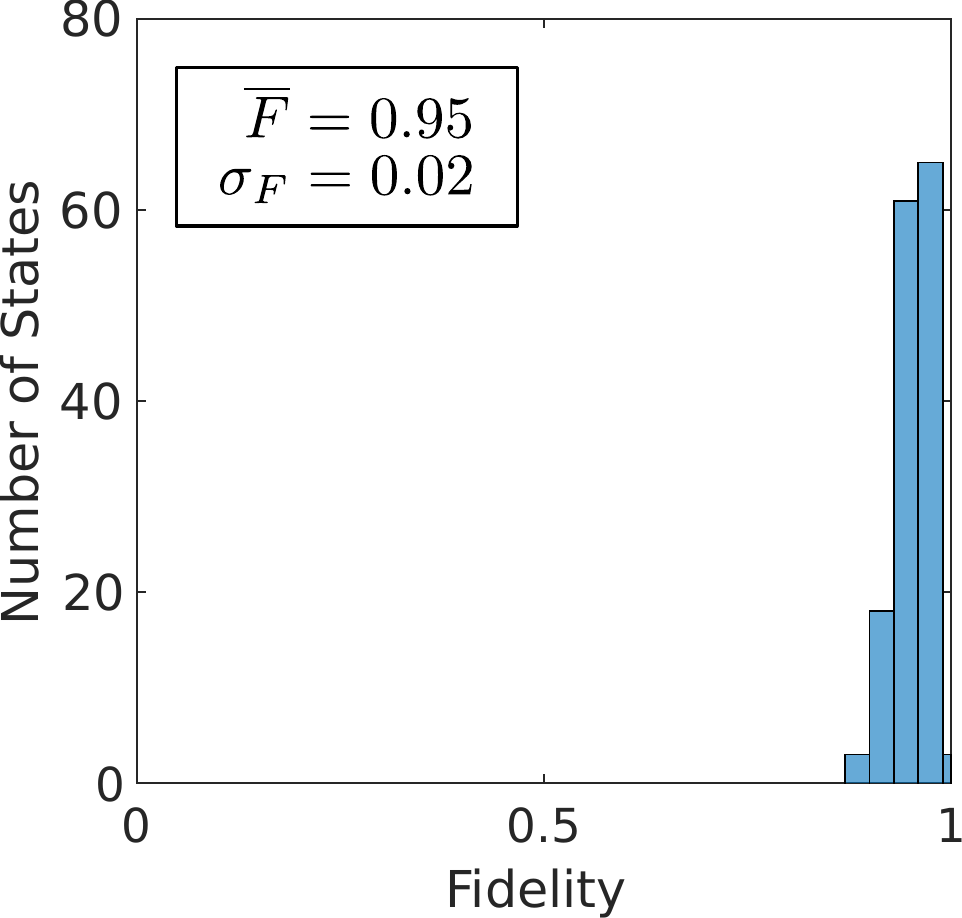}
  \caption{Histogram of the reconstruction fidelities for 150 random
    pure states of dimension $d=6$. The mean fidelity is
    $\overline{F} = 0.95$, and the standard deviation is
    $\sigma_F= 0.02$.\label{fig:histogram}}
\end{figure}

To assess the feasibility of the method we reconstructed a large
number of pure qudits of dimension $d=6$ randomly chosen.  As a figure
of merit we used the fidelity, that is defined, for pure states, as
$F(\Phi, \Psi) = |\langle\Phi|\Psi\rangle|$, where $|\Phi\rangle$
represents the state to be prepared, and $|\Psi\rangle$ the state that
is reconstructed \cite{Nielsen}. Ideally, $F=1$. The states were
selected with the Haar measure in a Hilbert space of dimension
$d=6$. The histogram in figure \ref{fig:histogram} shows the
occurrence fidelity for $150$ states. The mean fidelity is
$\overline{F} = 0.95$, with a standard deviation of $\sigma_F=
0.02$. This fidelity value is comparable to those obtained by means of
other QST methods \cite{Goyeneche2015,PearsStefano2019}, with the
advantage of being implemented from a very stable optical
architecture, which also allows to carry out the characterization
through only 4 measurements, regardless of the dimension of the
system.

\subsection{Turbulence correction}
\label{subsec:turbulence}
Since the PDI allows us to know the phase distribution of each point
of the wavefront, it is possible to evaluate deviations from the ideal
situation in the region outside the slits.  As this region does not
carry information about the states, its phase distribution should be
constant and known \emph{a priori} by the receiver. Then, the obtained
information about the aberrations present in this region can be used
to estimate and correct aberrations present in the ROIs of the
wavefront. In particular, we are going to deal with a case of great
interest in free-space quantum communications, such as the influence
of a turbulent medium on the information encoded and transmitted in
the quantum state of photons.

In a free-space communication system the sender encodes the
information in some degree of freedom of the wavefront, the signal
travels a given distance through free space, and is collected by means
of a telescope-like system by the receiver. The turbulence, which
entails random changes of the local pressure, and hence in the
refraction index, distorts the received signal. In the \emph{phase
  screen approximation} \cite{Boyd2011}, valid when the turbulence is
not too strong \cite{Rodenburg2011}, all the effects of the turbulence
can be represented as a random additive phase on the input of the
receiver's telescope. 
\begin{figure}
  \centering
  {\includegraphics[width=\linewidth]{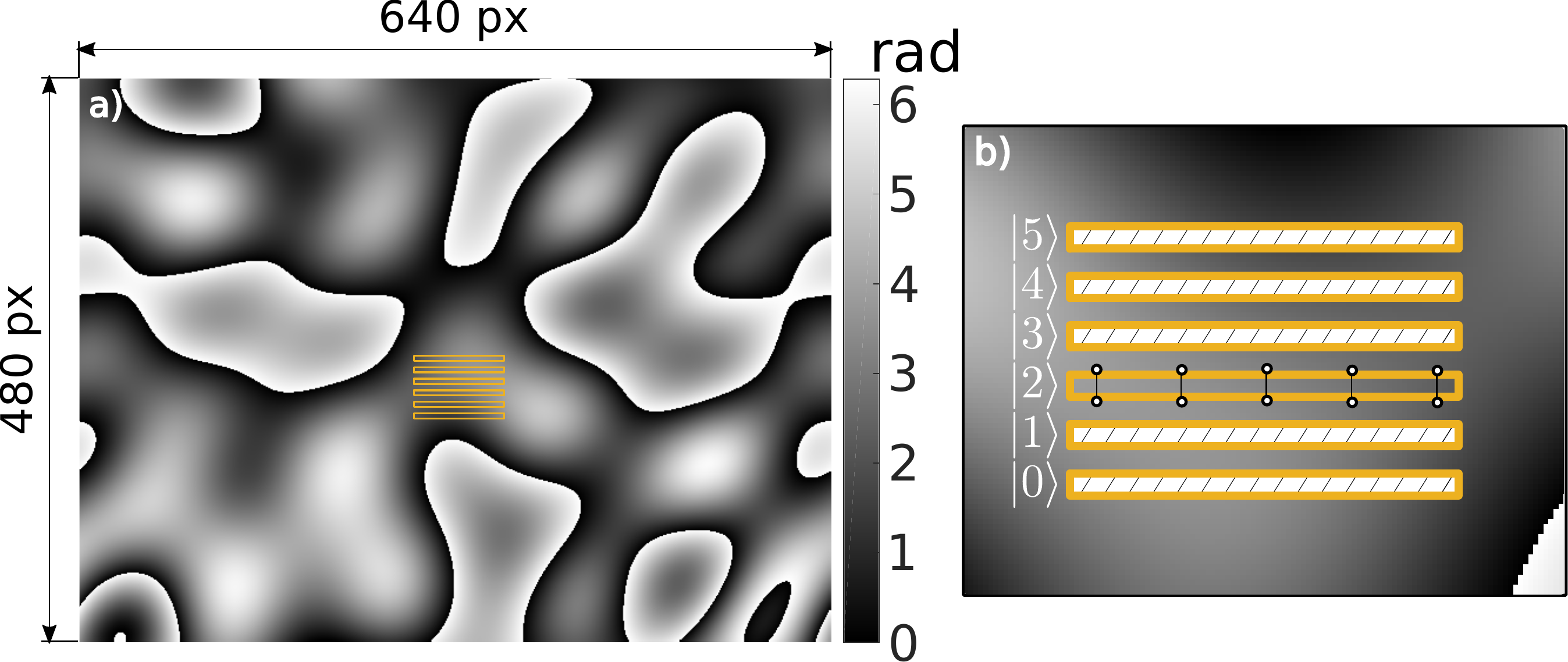}
  \phantomsubcaption\label{fig:mask-example:a}
  \phantomsubcaption\label{fig:mask-example:b}}
\caption{(a) Example of a turbulence phase mask displayed in the same
  modulator ($\mathbf{SLM_1}$) used to generate the slit states. This
  mask corresponds to an ensemble with Fried's parameter
  $r_0= 1.9\ \mathrm{mm}$. The yellow rectangles represent the
  position of the slits. (b) Schematics for the aberration estimation
  procedure inside the slit. Given that the phase aberration can only
  be evaluated on the background region, the phase for every point of
  the slit is obtained by interpolating the values nearest to the
  horizontal borders. The concept is illustrated for the slit
  $|2\rangle$.\label{fig:mask-example}}
\end{figure}
The stochastic set of phase masks is described by the Kolmogorov
theory, and follow the spatial structure function \cite{Fried1966}
\begin{align}\label{eq:dphi}
  \mathcal{D}_{\varphi} &= \langle\left[\varphi(\mathbf{r}) -
  \varphi(\mathbf{r}+\delta\mathbf{r})\right]^2\rangle \\\nonumber
  &= 6.88 (\delta r/r_0)^{5/3},
\end{align}
where $\langle \rangle$ denotes the average over the random ensemble,
$\mathbf{r}$ and $\mathbf{r} + \delta\mathbf{r}$ represent the
position of two points in the transverse plane of the telescope input,
and and $r_0$ is the Fried's parameter, that represents the typical
length of spatial correlation of the phase fluctuation.

In order to study the effect of turbulence over the propagation of our
slit states, an ensemble of phase masks verifying equation
\eqref{eq:dphi} was generated as a linear combination of normal modes
with random phases that obey an appropriate power law
\cite{Varga2018,Varga2018b}. The amplitude of the $m$ normal mode was
selected to be $A_m=A_0 e^{-1/4 m}$, where $A_0$ is the amplitude of
the fundamental mode. The phase of each mode is randomly changed in a
time $T_m =T_0 e^{-1/3m}$, where $T_0$ is the time for the fundamental
mode. The power law variations of the mode parameters ensure the
appropriate phase structure relation. As an example, one of the random
phase mask of this ensemble is shown in figure
\ref{fig:mask-example:a}.%
The empirical model of Hufnagel-Valley \cite{Valley1980} allows us to
relate the Fried's parameter with the height above the sea level in
which the transmission takes place. In our case, we have chosen
$r_0= 1.9\ \mathrm{mm}$, which is compatible with a $500\ \mathrm{m}$
long link at $647\ \mathrm{m}$.  Thus, to simulate a quantum state
reconstruction in presence of turbulence, we programmed, in the
$\mathbf{SLM_1}$, the nominal state to be reconstructed with the
addition of a random phase mask compatible with equation
\eqref{eq:dphi}. Then the four-step PSI is performed assuming that the
turbulent-induced phase aberration does not change during the
acquisition.  The validity of this assumption will depend both on the
acquisition rates of the camera and commutation time of the phase in
the PDI filter. Although in our case this time is limited by the
refresh rate of the LCD ($60\ \mathrm{Hz}$), there are alternative
electro-optical PDI filters with refresh rates up to $\mathrm{MHz}$
\cite{Paturzo2007}, and sCMOS cameras can exceed $1000\ \mathrm{fps}$
in small ROIs.

Let us describe the method to correct the phase aberrations introduced
by the turbulence. As we explained in section \ref{sec:method}, the
background signal added to the codification of our photonic states
allows us to reconstruct the phase distribution of the whole region
within the circular pupil.  Thus, we can accurately estimate the
turbulence-induced phase distortion \emph{outside} the ROIs. Inside
the ROIs we are unable to discriminate the phase aberrations from the
unknown phase of the slit, and then, the turbulence-induced phase
distortion cannot be estimated directly. However, the information of
the phase \emph{outside} the ROI can be used to interpolate it inside
of each slit. Figure \ref{fig:mask-example:b} exemplify this idea for
the slit assigned to the state $|2\rangle$. Along the slit, the phase
values corresponding to the points in the circles (outside the ROI)
are used to calculate the phase values of the points belonging to the
straight line (inside the ROI) by means of a linear interpolation. In
that way, the full map of the phase aberrations can be reconstructed
and subtracted from the full phase distribution obtained by PSI.

\begin{figure}
  \centering
  {\includegraphics[width=0.9605\linewidth]{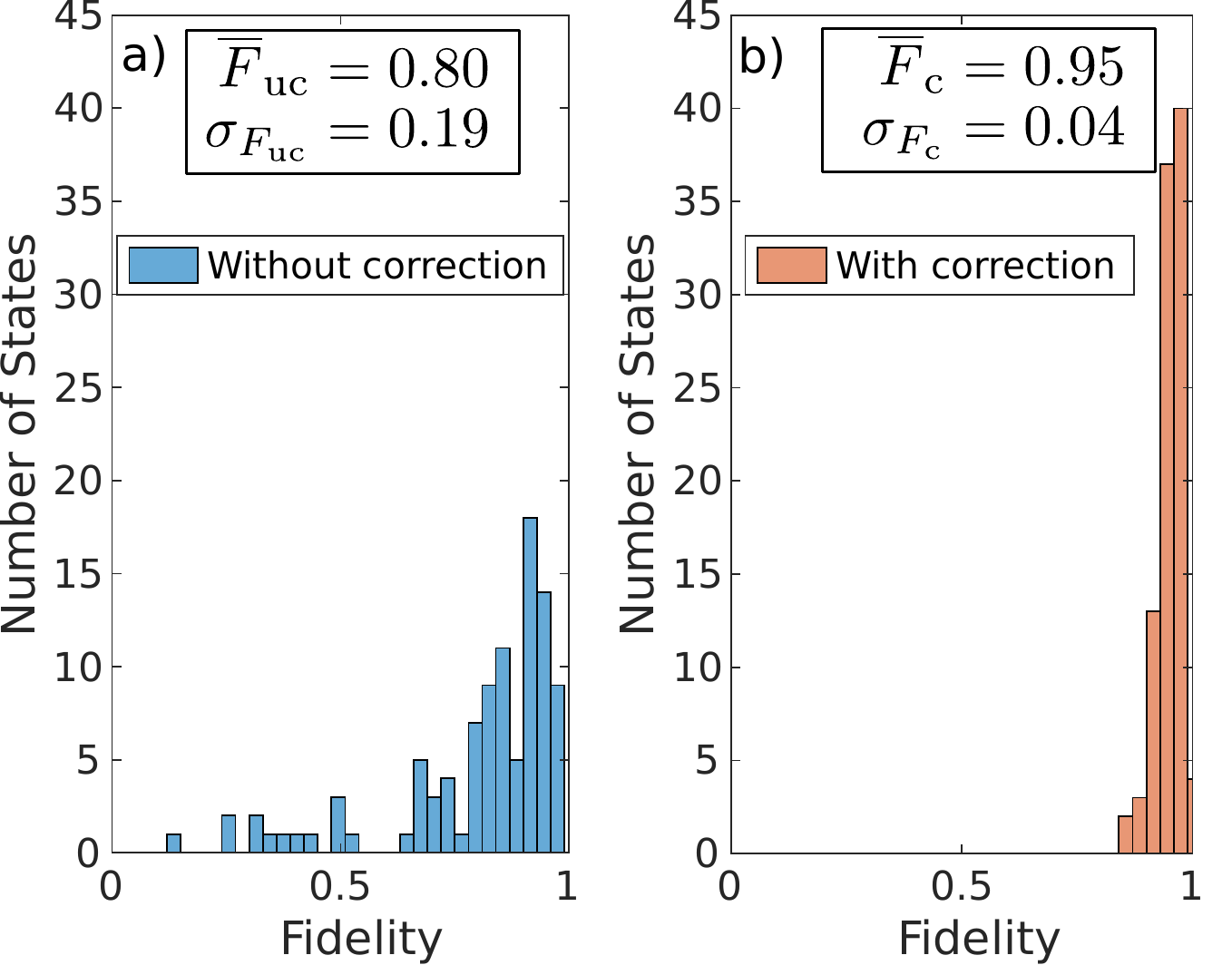}
  \phantomsubcaption\label{fig:hist-turb:a}
  \phantomsubcaption\label{fig:hist-turb:b}}
  \caption{(a) Histogram of fidelity reconstruction for 100 random
    states ($d=6$) affected by a turbulence with
    $r_0 = 1.9\ \mathrm{mm}$. Its mean fidelity is low,
    $\overline{F_{\mathrm{uc}}} = 0.8$. (b) Histogram of the reconstruction
    fidelity for the same 100 states \emph{after} applying the phase
    correction method. The mean fidelity
    $\overline{F_{\mathrm{c}}} = 0.95$ is comparable to the case
    without turbulence. \label{fig:hist-turbulence}}
\end{figure}

To test the capability of the method to correct the turbulence
aberrations we have performed the reconstruction of $100$ random
states of dimension $d=6$. Figure \ref{fig:hist-turb:a} shows the
histogram for the reconstruction fidelity when the effects of the
turbulence turbulence are not corrected. A significantly lower mean
fidelity ($\overline{F_{\mathrm{uc}}} = 0.8$) than in the case without
turbulence, and a higher standard deviation
($\sigma_{F_{\mathrm{uc}}} = 0.19$), are
obtained. Figure \ref{fig:hist-turb:b} shows the occurrence of
reconstruction fidelity for the same states after being processed with
the described correction method. We can say that there is an excellent
enhancement in the quality of the reconstruction supported by a mean
fidelity of $\overline{F_{\mathrm{c}}} = 0.95$ and a standard
deviation $\sigma_{F_{\mathrm{c}}}= 0.03$, both values comparable to
those of the situation without turbulence.

\section{Conclusions}
\label{sec:conclusion}
We have presented a method to reconstruct pure spatial qudits codified
in the discretized transverse momentum of a photon field (usually
called slit states), that is based on a point diffraction
interferometer. This architecture results in an experimental setup
easy to align, and very stable against vibrations and air
turbulence. As remarkable property, the measurements can be
parallelized by using a multipixel detector,in which case only four
interferograms are required to reconstruct any pure state regardless
of the dimension of the system. To this end we propose an alternative
encoding for the slit states that includes an uniform background that
plays the role of reference beam. This allows to accurately
reconstruct the phase of the whole photonic wavefront, and estimate
possible aberrations.

We have experimentally tested the proposed method in the
reconstruction of states of dimension d=6, obtaining a mean fidelity
value $\overline{F}=0.95$. Finally, we presented a proof-of-principle
demonstration that the method can be used to to correct the influence
of turbulence in a free-space communication. To that end, we
experimentally simulated a turbulent channel using random phase
masks. In the case of weak turbulence, and after correction, we
recovered mean fidelity values comparable to the case without
turbulence.